\documentclass[preprint]{revtex4}

\usepackage{graphicx}
\def\underlinewords#1{%
	\def\stuff{#1 }\leavevmode\expandafter\ulword\stuff * }
	\def\ulword#1 {\def\one{#1} \ifx\one\aster\let\next\relax
	\else\vtop{\hbox{\strut#1}\hrule\relax}
	\let\next\ulword\fi\next}
\def\strikeoutwords#1{%
	\def\stuff{#1 }\leavevmode\expandafter\soword\stuff * }
	\def\soword#1 {\def\one{#1} \ifx\one\aster\let\next\relax
	\else\vtop{\hbox{\strut#1}\kern-.5\baselineskip\hrule\relax}
	\let\next\soword\fi\next}
\def\aster{*}
\begin{document}

\preprint{ANL-HEP-CP-02-13}

\title{Future Possibilities for Lepton-Hadron Collider Physics and Detectors}


\author{G. Fleming}
%\thanks{}
\affiliation{Ohio State University, Columbus, Ohio}

\author{E. Kinney}
\affiliation{University of Colorado, Boulder, Colorado}

\author{S. Lammers}
\affiliation{University of Wisconsin, Madison, Wisconsin}

\author{S. Magill}
\affiliation{Argonne National Laboratory, Argonne, Illinois}


\date{\today}

\begin{abstract}
We have considered the physics opportunities of future lepton-hadron colliders
and how these opportunities might be realized in a possible polarized
eRHIC facility 
and an e-p collider as part of a staged or final version VLHC.  We evaluated
the physics priorities based on experience at HERA and, using simulated
data for 
e-p collisions with $\sqrt{s} > 1$~TeV, showed how detector designs would be
impacted by the physics.
\end{abstract}

% \pacs{}
%\keywords{}

\maketitle


\section{Introduction}
The Deep Inelastic Scattering (DIS) process, in which an electron (or
any lepton) 
is used to destructively probe a hadron has proven to be one of the best 
ways to study hadronic structure and the nature of the strong
interactions between its  
fundamental constituents (quarks and gluons).  Over the past 30 years,
the theory of 
color interactions of quarks and gluons, Quantum Chromodynamics (QCD), has been
validated to distances at least 10 times smaller than the size of the
proton.  A future 
e-p collider would have the following attractive advantages :

\begin{itemize}
\item The (pointlike) electron probe is ideal for examining proton structure.
\item It is tunable over a wide range in hard scattering scale.
\item QCD parameters and properties can be studied in a relatively
background free  
environment.
\item Some unique couplings may exist to new physics at high energy.
\end{itemize}

The first three points, taken together, define an e-p collider as a "QCD
Factory" where 
all aspects of the strong interaction can be investigated, including short- to 
long-distance interactions between quarks and gluons, low- to high-density 
partonic states, and detailed measurements of the proton structure and its 
evolution.

As suggested by our subgroup charge, we considered and evaluated two
very different 
future e-p options - a low energy collider with the capability to
provide e-A collisions
as well as polarized e-p interactions, and a high energy e-p machine
with $\sqrt{s} >  
1$~TeV.  

\section{Electron-Ion Collider (EIC)}
At two recent Town Meetings held at JLAB in December~2000 and at BNL in
January~2001, it 
was recommended to the Nuclear Sciences Advisory Committee (NSAC) that a
high-luminosity electron-ion collider covering CM energies in the range
of $30-100$~GeV be built as the next generation facility for the study
of electromagnetic and
hadronic physics~\cite{eicw}.  The purpose of this collider would be to
answer the following questions:
\begin{itemize}
\item What is the structure, both flavor and spin, of hadrons in terms
of their quark 
and gluon constituents?
\item How do quarks and gluons evolve into hadrons via the dynamics of
confinement? 
\item How do quarks and gluons reveal themselves in the structure of
atomic nuclei? 
\end{itemize}

In addition to these questions, the Electron-Ion Collider would
contribute to a fundamental
understanding of QCD by testing its predictions and parameters in
extreme conditions and
at the limits of its expected applicabilities.  In particular, parton 
distribution functions (pdfs) could
be measured in regions of parameter space in which non-linear effects
are expected to 
dominate.  Tests of QCD in regions where many-body effects between
strongly-interacting 
matter are expected could be done.  An even more fundamental question is
whether nuclei 
can be used to study partonic matter under extreme conditions.

\subsection{EIC Parameters}
The EIC should be built with the following characteristics :
\begin{itemize}
\item Capable of collisions between electrons (positrons) and protons,
light and heavy 
nuclei
\item Provide high luminosity $L \ge 10^{33}\,{\rm cm^{-2}\,s^{-1}}$ per
nucleon 
\item Operate in a wide range of CM energies $E_{cm} = 15 \; {\rm GeV
\rightarrow 100 \; GeV}$
\item Polarization of electron and proton spins
\item Two interaction regions with dedicated detectors
\end{itemize}

%Fig.~\ref{fig:eic} shows the kinematic region in $x-Q^2$ space accessible at the EIC
%compared to past fixed target experiments and to the present HERA ep collider (no proton
%polarization or light and heavy ions).
%\begin{figure}[hbt]
%\includegraphics{fig_1}
%\caption{\label{fig:eic} Kinematics at EIC compared to HERA and fixed-target experiments.}
%\end{figure}

\subsection{EIC Physics Priorities}
The following list of physics topics related to QCD could be addressed
at the EIC : 

{\bf Flavor and Spin Structure of the Nucleon :} It would be possible to
measure the pdfs
of light quarks and the gluon by tagging final states in inclusive DIS.
For example, by
tagging kaons, both momentum and spin pdfs of strange quarks could be 
determined with high
precision down to $x \simeq 10^{-3}$.  In general, the spin structure of
the nucleon could
be determined to lower $x$ than is presently known, leading to a better
understanding of the contribution of the sea to the total spin of the
nucleon.  Finally,
a complete understanding of the partonic structure of a nucleon requires
not only 
momentum and spin contributions, but also those from, for example,
parton-parton 
correlations.  These fall into a class of structure functions called
Generalized Parton 
Distributions, many of which could be accessed at the EIC.   

{\bf Partonic Substructure of Mesons and Hyperons :} Very little is
known about the
structure of hyperons and mesons, despite their role as the glue that holds
nuclei together.  At the EIC, measurements of the quark and gluon
structure of these particles could be compared to those in the nucleon.
Since mesons and
hyperons are the Goldstone bosons of spontaneously broken chiral
symmetry, fundamental 
questions of the role of quarks and gluons in transition from partonic
degrees of  
freedom to their Goldstone modes could also be addressed.

{\bf Hadronization :}  At the EIC, complete final states of the hard 
scattering process
could be detected and reconstructed, allowing studies to be done of the 
transformation by
colored quarks and gluons into color neutral hadrons.  This
non-perturbative process must
be measured experimentally in order to understand the long-range
dynamics of confinement.
Using flavor tagging and jet reconstruction, studies of the transfer of
spin from quarks
to hadronic final states could also be done.

{\bf Role of Quarks and Gluons in Nuclei :}  Comparisons of collisions 
involving both
light and heavy nuclei at the EIC could illustrate possible effects of 
nuclear matter on the
parton distributions, for example, the effects of nuclear binding,
expressed as exchanged
mesons, on the underlying quarks and gluons in nucleons.  The large
range in $x$ at the EIC
allows the $x$-dependence of these and other nuclear effects to be
determined.  Also, the
initial conditions for relativistic heavy ion collisions could be
determined, helping to 
fill in missing pieces of the transition from cold, partonic matter to a
hot, dense 
quark-gluon plasma.

{\bf High Density QCD at Low $x$ :}  At HERA, the scaling dependence of
$F_2$ at low $x$ 
indicates that the gluon density increases very steeply as $x$ decreases.
Unitarity bounds
on the gluon distribution
predict that this steep rise must eventually saturate, at around $x
\simeq 10^{-6}$ for 
$Q^2 = 10 \; {\rm GeV^2}$.  Across this boundary, non-linear evolution of
parton distributions 
occurs, leading to a new state of color-saturated high-density QCD.
Since the parton 
densities in nuclei are enhanced by a factor of $A^{1/3}$, these effects
might be seen in
nuclei at lower energies than in protons.  The EIC could  be the first
place to see this
new state of partonic matter, if nuclear effects are negligible or
non-existent in the 
low $x$ region. 

\section{ep at $\sqrt s >$ 1 TeV}
There are presently no e-p colliders planned at $\sqrt s > 1$~TeV.  A
possible upgrade  
to HERA, THERA~\cite{tbook}, in which 250~GeV electrons from
TESLA~\cite{tdr} are collided 
with 900 GeV protons from HERA, has been studied.  This option (at
$\sqrt s \sim 1$~TeV) 
could only be realized if TESLA is built at DESY, if the accelerators are made 
to match at an appropriate IP, and if luminosity is taken from the TESLA
program  
and provided for THERA.  

At the Snowmass 2001 workshop, studies were made of an e-p option in
which electrons 
from a future $e^+ e^-$ linear collider are collided with protons from a
future  
VLHC \cite{vlhc}.  
Our "standard" option was defined as a 250~GeV $e^-$ interacting with a
20~TeV proton,  
corresponding to Stage I of the VLHC (epVLHC).  The corresponding CM
reach of this option 
is $\sqrt{s} \sim 4.5$~TeV.  The ratio of incoming beam energies is
$E_p/E_e = 80$, more
than twice the asymmetry at HERA.  As an alternative, one of the THERA options,
800~GeV $e\times 800$~GeV $p$ was studied as an example of a symmetric
collider.  Both the 
asymmetry of the beams and the physics requirements affect the design of
detectors at a 
future e-p facility.

\subsection{Physics Priorities}
In the early planning for HERA, it was thought that new physics beyond
the Standard  
Model was within reach.  HERA would see at least some of the following:
leptoquarks,  
excited electrons, contact interactions indicating quark substructure,
SUSY, and  
others.  Now we know that the Standard Model is alive and well at HERA.
However, 
HERA has expanded our knowledge of the proton structure by several orders of 
magnitude in both hard scattering scale, $Q^2$, and longitudinal
momentum fraction, $x$.
It has shown that the inclusive proton structure function, $F_2$, rises
steeply as $x$  
decreases.  This steep rise has been linked to a large increase in the gluon 
distribution in the proton at low $x$.  HERA has also discovered that
hard diffractive  
scattering contributes a significant amount to the total DIS cross
section.  Also,  
at HERA, extensive inclusive and exclusive measurements have been made
to test the  
limits of perturbative QCD and to study the transition to the non-perturbative 
regime.

So, we have learned that e-p at HERA is an ideal place to study the
dynamics of  
perturbative QCD.  We have, therefore, taken this lesson from HERA and applied 
it to our determination of physics priorities at a future e-p collider.
We believe  
that the following ordered physics priorities should form the basis of
the physics  
case for a future e-p collider.  

\subsubsection{Perturbative QCD Dynamics at Low $x$}
The evolution of partons in the proton is normally described by a set of linear
equations where the number of partons evolves as a function of either
$Q^2$ (DGLAP 
evolution), $x$ (BFKL evolution), or a combination of both $Q^2$ and $x$
evolution 
(DLLA).  The following table summarizes the forms of linear parton evolution:
\begin{table}[hbt]
\centering
\begin{tabular}{|l|c|c|c|} \hline
 & DGLAP & DLLA & BFKL \\ \hline
Resummed terms & $(\alpha_s \ln Q^2)^n$ & $(\alpha_s \ln Q^2 \ln 1/x)^n$ & 
$(\alpha_s \ln 1/x)^n$ \\  \hline
Region & High/medium $x$ & low $x$ & small $x$ limit \\ \hline
Program & Fixed target DIS & HERA/THERA & epVLHC \\ \hline 
\end{tabular}
\end{table}

Outstanding questions remain about the form of the linear evolution
processes seen  
in the HERA DIS data.  At HERA, DGLAP alone can describe the full range
of inclusive 
$F_2$ data as long as enough gluons are included in the proton at a low
scale.   
Some more exclusive measurements, e.g., final state multiplicities in
the current 
region of the Breit frame, are consistent with DLLA evolution, and from
forward jets 
and hadrons, it also appears that DGLAP alone is unable to describe the data.
The tunability of the electron probe is a distinct advantage in
measurements of this 
type, since it allows for selection of extreme kinematic regions where one 
evolution scheme is expected to dominate over the others.  

A future e-p collider at very high energy could access the very small
$x$ region where 
BFKL evolution should apply.  At some point, DGLAP should fail to
describe the data 
when large
longitudinal distances are being probed (very small $x$).  As the number
of partons 
increases, which is evident at HERA with no signs of stopping,
eventually parton-parton 
interactions should start to occur.  The eventual trade-off between
parton evolution and
parton-parton recombination "saturates" the cross section and the steep rise
becomes flat.  The complete description of parton evolution would then
require the addition of non-linear terms.  This should occur when the
number of gluons 
is $xg \sim Q^2/ \alpha_s(Q^2)$.  
HERA has had much success in mapping the low-$x$ growth in $F_2$, but
has not yet seen  
effects that could be interpreted as saturation of the proton structure 
function.  Extrapolating with GRV pdfs, at $Q^2 = 10 \; {\rm GeV^2}$,
the saturation boundary would occur at $x \simeq 10^{-6}$.  

\begin{figure}
\includegraphics[width=5.in]{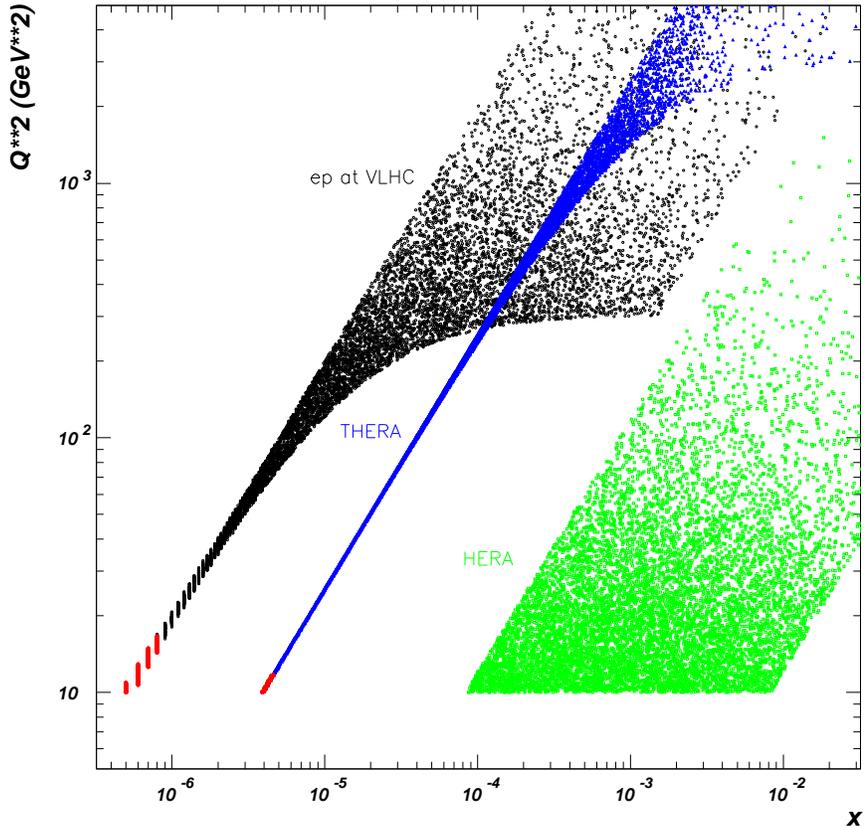}
\caption{\label{fig:hdqcd} Kinematics at HERA, THERA, and epVLHC,
indicating where saturation
is expected to occur.}
\end{figure}

At Snowmass, simulated events were generated using the CASCADE Monte
Carlo program~\cite{hjcas}
which incorporates the CCFM parton evolution model. This model should be
a good choice at
a future e-p collider since it allows for $x$-evolution of the partons
(BFKL) in the small
$x$ limit.  Fig.~\ref{fig:hdqcd} shows the e-p scattering 
kinematic plane in $x$ and $Q^2$ comparing HERA with THERA and epVLHC.  

A minimum scattering angle for the electron of $10^\circ$ was required
for each option 
and shows up as the high $x$, low $Q^2$ limit of the events generated
for THERA and epVLHC.
The low $x$, high $Q^2$ limit is the kinematic constraint, $y \leq 1$.  The red
shaded region denotes the part of the region where $\ln \; 1/x$ is at
least a factor of 5
larger than $\ln \; Q^2$.  For $Q^2 = 10 \; {\rm GeV^2}$, this is where
saturation effects should
occur.

\begin{figure}[hbt]
\includegraphics[width=4.in]{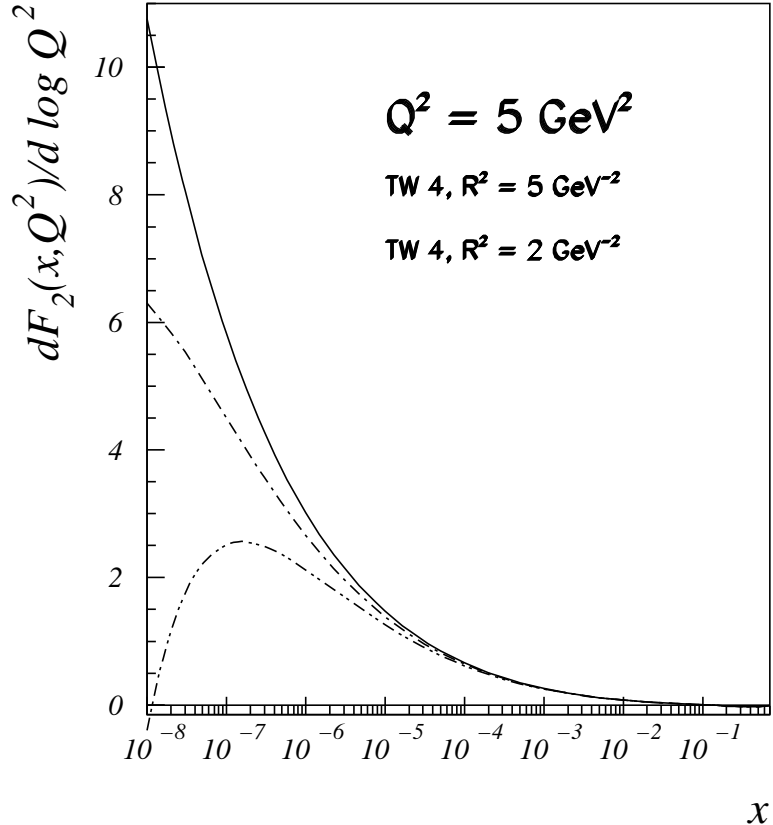}
\caption{
$F_2$ slope $dF_2(x,Q^2)/d\log Q^2$ at $Q^2 = 5~{\rm GeV^2}$.  Full line:
leading order twist-2;
dash-dotted line: twist-4 contribution with a screening
radius $R^2 = 5~{\rm GeV^{-2}}$; dash-dot-dot line: $R^2 = 2~{\rm GeV^{-2}}$.}
\label{fig:twist}
\end{figure}

The onset of saturation is also 
affected by the twist level (higher order $(1/Q^2)^n$ terms of the
evolution equations). 
Current measurements of $F_2$ are well-described by twist-2 evolution
equations,  
but even after considering higher-order corrections these equations do not 
lead to saturation as $x \rightarrow 0$.  However, with appropriate
choices of pdfs and 
screening radius (size of the region in the proton in which
parton-parton interactions 
begin to occur), twist-4 contributions to the evolution equations can 
influence the onset
of saturation effects.   
One particular parameterization (CTEQ) shown 
in Fig.~\ref{fig:twist} indicates that with an appropriate choice of
screening radius,  
the inclusion of twist-4 contributions predicts the onset of 
saturation at around $x=10^{-7}$.   

At even lower $x$, beyond the saturation boundary, perturbation theory
breaks down even 
if $\alpha_s$ is small, because the non-linear effects are large.  The
physical state 
of quarks and gluons in this region has been described as a Color-Glass
Condensate,  
named from the type of equations used to describe high-density
concentrations of the 
colored gluons.  If this state exists, it should be seen at a future e-p
collider with 
$\sqrt{s}$ in the few TeV range.

\subsubsection{Proton Structure and Flavor Decomposition}
In addition to measurements of inclusive proton structure functions, an
e-p collider is 
also an ideal place to decompose the inclusive $F_2$ measurement into
individual flavor 
pdfs. In searches for new physics at p-p machines, e.g., at the LHC and 
VLHC, it will be 
important to know the contribution to cross section measurements from the heavy
quark content of the proton. As is the case today, all measurements of
the $b$-quark cross
section, including those at HERA, are larger than the predicted
theoretical values, and 
without a precise measure of the $b$-content of the proton, this
difference is hard to  
interpret as new physics. 

\begin{figure}[hbt]
\includegraphics[width=5.5in]{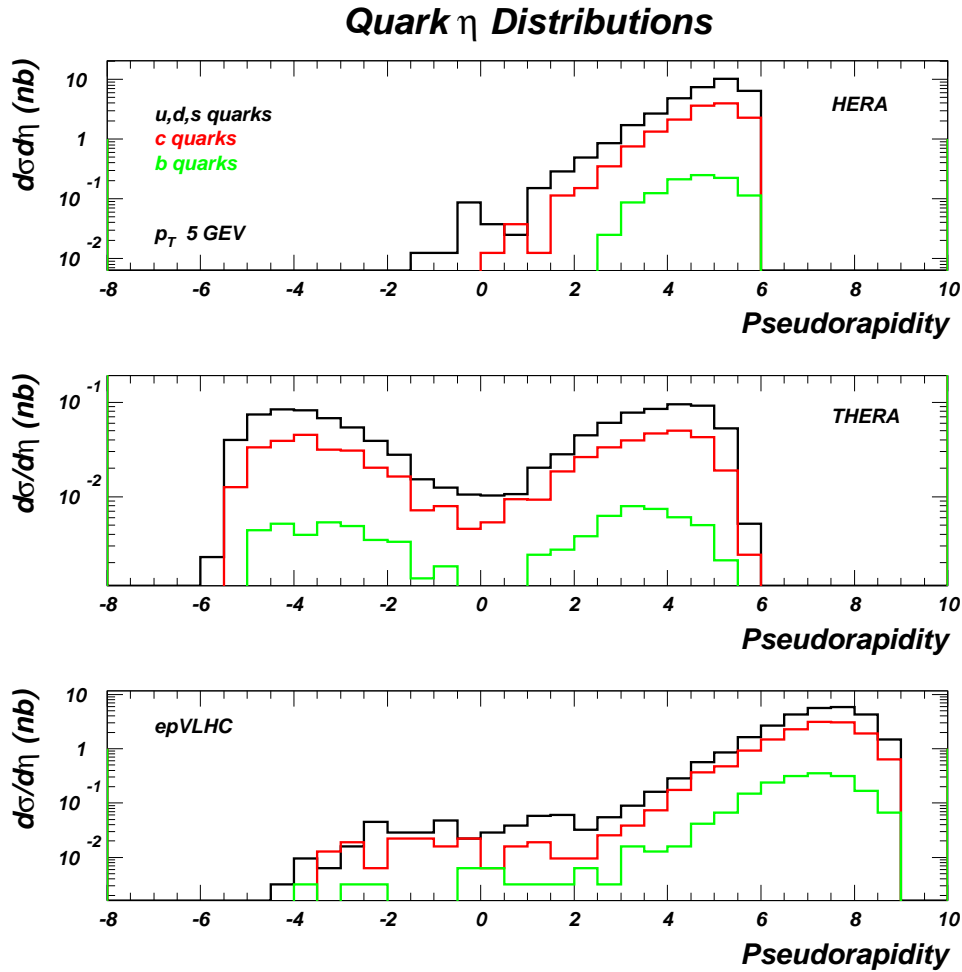}
\caption{\label{fig:hqeta} $\eta$ distributions of quarks at HERA,
THERA, and epVLHC.} 
\end{figure}

Comparisons between HERA, THERA, and epVLHC for heavy quark distributions were
examined in MC simulated events.  Fig.~\ref{fig:hqeta} shows the light
and heavy quark  
distributions in pseudorapidity ($\eta$) for the three e-p options.  

A minimum $p_T$ requirement of
$6 \; {\rm GeV}$ was imposed on the quarks.  For both HERA and epVLHC,
the quarks are forced into
the forward (proton direction) region, requiring a detector which would 
look more like a
fixed target detector than a collider detector. However, for the symmetric
case of THERA, the quarks are scattered closer to the central region. A
``standard'' 
symmetric collider detector with vertex detector and some extended
coverage in the 
forward region might work for this configuration.  

\subsubsection{Beyond the Standard Model (BSM)}
Although the opportunities for studying exotic physics at a future e-p
facility are  
potentially rich and varied, we did not attempt any predictive studies
at Snowmass.  
However, the following discussion summarizes the current state of some 
BSM investigations 
and indicates where a future e-p collider could contribute.

{\bf Leptoquarks (LQ) and squarks :} An e-p collider is ideal for the
production of 
new bosons possessing both leptonic and baryonic quantum numbers, e.g., 
leptoquarks and 
squarks in SUSY with R-parity violation.  HERA has set limits on the
production of these
particles that are competitive with comparable limits set at the
Tevatron.  If LQs 
are discovered in a hadron-hadron collider (e.g. LHC), a future e-p
collider would be the 
best place to study their properties: the angular distribution of the 
final-state lepton can discriminate between scalar and vector
resonances; the LQ fermion
number is determined by comparing the signal cross section in $e^+$p and $e^-$p
collisions; the chirality structure
of the LQ coupling can be determined by the polarization of the incoming
lepton beam.   
Further constraints on the LQ coupling could be made by polarizing the 
proton beam as well. 
At an e-p facility at the VLHC, s-channel LQ resonances in the mass
range up to  
$\sim 4.5 \; {\rm TeV}$ could be seen directly, as well as higher mass states
by parametrization
as a contact interaction.

{\bf Contact Interactions :} In addition to very high mass LQs, generic
four-fermion 
contact interactions could also signal new physics processes, e.g., by 
interfering with the 
observed NC DIS cross section.  If contact interactions were seen, e-p 
collisions would 
give a unique insight into the chiral structure of the interactions by
exploiting the  
lepton beam polarization.  

{\bf Excited Leptons :} Although the LHC will be able to discover $e^*$ 
and $\nu^*$ by
pair production independent of coupling, some processes involving these
particles are  
better done at an e-p collider.  For example, searches for $\nu^*$ could
take advantage 
of the large u-quark density in the proton and the helicity nature of the 
charged-current interaction.   

{\bf Large Extra Dimensions :} The $t$-channel exchange of Kaluza-Klein 
gravitons in models 
with large extra dimensions affects the $Q^2$ distribution of NC DIS
events.  While the LHC
again can probe very large compactification scales, some models predict
that fermions with
different gauge quantum numbers are localized on different branes.  A
future e-p 
machine could provide complementary information on fermion localization,
since the 
two-gluon final state would dominate the cross section at the LHC.

\subsection{Detector Considerations}
Since it is easier to accelerate protons to higher energy than electrons
and positrons, 
the highest $\sqrt{s}$ e-p machines would be highly asymmetric in
incoming beam energies.
At HERA, the proton beam energy is $\sim 30 \times$ that of the lepton
beam.  At the VLHC,
with $20 \; {\rm TeV}$ protons colliding with $e^+ e^-$ linear collider
electrons of  
$250 \; {\rm GeV}$,
the asymmetry factor is $\sim 80$.  Therefore, the detector must be
highly asymmetric 
with most of the detection capability far forward in the proton direction.  The
detection of scattered
electrons would also require calorimetry in the opposite direction to
the proton.  For 
studies of QCD dynamics and non-linear evolution effects, an asymmetric
detector 
would suffice, including good angle and energy measurement of the
scattered electron 
and the capability of hadron/jet reconstruction in the far forward region.  
Since the objective
is an inclusive measurement at extremely low $x$, a split detector of
this type would be
necessary at such a highly asymmetric collider facility.

For measurements of flavor pdfs, a more symmetric configuration would be
desirable, 
however, as already mentioned, this would come at the expense of
$\sqrt{s}$.  Even with 
symmetric beam energies, the heavy quarks populate the forward region of
the detector, 
so forward detection is still important.  Detectors for more symmetric
beams look more 
like traditional collider detectors.

For studies of possible BSM processes, it is important to have a
hermetic $4\pi$ detector. 
This is very difficult to do with highly asymmetric beams, so specific
BSM searches would
be chosen at the detector design stage.  This limits the effectiveness
of a future e-p 
machine as a BSM search device, hence our lower priority for this type
of physics. 

\section{Conclusions}
We have attempted to understand the important physics issues for
possible future e-p 
colliders.  Both the low $\sqrt{s}$ (EIC) and high energy options
(epVLHC) are seen 
primarily as QCD "Factories" which utilize the advantages of a
pointlike, well-defined 
probe to measure the parameters and dynamics of QCD theory.  Our aim was
to provide 
a general, first view of future e-p options, so that if, at some future
date, an e-p 
collider configuration becomes viable, specific physics and detector
studies could build
on our ideas as contained here.

\end{document}